\documentclass[fleqn,usenatbib]{mnras}

\usepackage{newtxtext,newtxmath}

\usepackage[T1]{fontenc}

\DeclareRobustCommand{\VAN}[3]{#2}
\let\VANthebibliography\thebibliography
\def\thebibliography{\DeclareRobustCommand{\VAN}[3]{##3}\VANthebibliography}

\usepackage{graphicx}	
\usepackage{amsmath}	
\usepackage{amssymb}	
\usepackage{xcolor}

\def\sw{Swift~J004427.3-734801}

\title[\sw]{\sw - a probable Be/white dwarf system in the Small Magellanic Cloud}

\author[M. J. Coe et al.]{
M.~J. Coe,$^{1}$\thanks{E-mail: mjcoe@soton.ac.uk}
J.~A. Kennea,$^{2}$, P.~A. Evans $^{3}$
and A. Udalski$^{4}$
\\
$^{1}$Physics \& Astronomy, The University of Southampton, SO17 1BJ, UK\\
$^{2}$Department of Astronomy and Astrophysics, The Pennsylvania State University, 525 Davey Lab, University Park, PA 16802, USA\\
$^{3}$University of Leicester, X-ray and Observational Astronomy Research Group, School of Physics \& Astronomy, University Road, Leicester LE1 7RH, UK\\
$^{4}$Astronomical Observatory, University of Warsaw, Al. Ujazdowskie 4, 00-478 Warszawa, Poland \\
}

\date{Accepted 2020 June 8. Received 2020 June 5; in original form 2020 April 28.}

\pubyear{2020}

\begin{document}
\label{firstpage}
\pagerange{\pageref{firstpage}--\pageref{lastpage}}
\maketitle

\begin{abstract}
\sw\ is an X-ray source in the Small Magellanic Cloud (SMC) that was first discovered as part of the Swift S-CUBED programme in January 2020. It was not detected in any of the previous 3 years worth of observations. The accurate positional determination from the X-ray data have permitted an optical counterpart to be identified which has the characteristics of an \textcolor{black}{O9V-B2III star. Evidence for the presence of an IR excess and significant I-band variability strongly suggest that this is an OBe type star.} Over 17 years worth of optical monitoring by the OGLE project reveal periods of time in which quasi-periodic optical flares occur at intervals of $\sim$21.5d. The X-ray data obtained from the S-CUBED project reveal a very soft spectrum, too soft to be that from accretion on to a neutron star or black hole. It is suggested here that \textcolor{black}{this a rarely identified} Be star-white dwarf binary in the SMC.

\end{abstract}




\section{Introduction}

It has been predicted for a while, from binary system evolutionary models, that there should be a large number of Be/white dwarf (BeWD) systems compared to the number of Be/neutron (BeNS) star systems - see, for example \cite{raguzova2001} who predicts 7 times more BeWD systems that BeNS systems. Despite this, and despite the identification of over 70 BeNS star systems in the SMC \citep{ck2015, haberl2012}, so far only \textcolor{black}{a small number of BeWD systems have been possibly identified in this galaxy  \citep{sturm2012, cracco2018}. Similarly, a few BeWD systems are proposed to exist in the Large Magellanic Cloud - for example, XMMUJ~052016.0-692505 \citep{k2006} and RX J0527.8-6954 \citep{oliveira2010}. In addition, several supersoft sources (SSS) have been identified in M31 as possible BeWD systems\citep{orio2010}. But, in total, extremely small numbers compared to the known BeNS populations and binary model predictions.}

It is possible that some observational constraints may be making such BeWD systems harder to detect, For example, the optical channel offers no obvious route to separating BeWD and BeNS systems. However, in the X-ray a \textcolor{black} {possible distinguishing features may be very high luminosity above $10^{36}$ erg/s in the soft X-ray range (below 0.8 keV)    indicating a nuclear burning white dwarf} - that of a very soft spectrum for a BeWD compared to that of a BeNS. Such SSS have been long identified as a characteristic of accretion and nuclear burning on the surface of a white dwarf (see, for example \cite{kahabka2006}). X-ray luminosities in the range $10^{35} - 10^{38}$ erg/s are predicted - well within the capabilities of current X-ray observatories. It is clearly very important for binary system evolution models that an accurate estimate is obtained of the the number of BeWD systems in the SMC.

In this paper we report on both X-ray and optical recent photometry of a newly identified system \sw.  The X-ray object was identified \citep{coe2020} as part of the S-CUBED regular monitoring of the SMC \citep{kennea2018} using the {\it Neil Gehrels Swift Observatory} \citep{gehrels04}. Interesting X-ray variability and, crucially, the presence of a soft X-ray spectrum suggests that this might be \textcolor{black}{one of very few} BeWD system identified in the SMC.  In addition historical optical photometric data are reported here from the OGLE project showing an occasional, quasi-periodic, rapid flaring behaviour of the counterpart over the last $\sim$17 years. This behaviour is almost certainly related to a binary period of ~21d.

Therefore it is proposed here that the optical counterpart to this X-ray object is an OB star and thus that this is a newly identified BeWD system. Only \textcolor{black}{one of a handful} in the Magellanic Clouds as a whole.

\section{Observations}

\subsection{S-CUBED detection and follow-up}

\sw\ was first detected by the S-CUBED survey \citep{kennea2018}, a shallow weekly survey of the optical extent of the SMC by the Swift X-ray Telescope (XRT; \citealt{burrows05}). Individual exposures in the S-CUBED survey are typically 60s long, and occur weekly, although interruptions can occur due to scheduling constraints. Starting with the S-CUBED observation taken on January 22nd, 2020, S-CUBED detected an previously uncatalogued X-ray source, internally numbered SC1764, at a count rate of $0.053 \pm 0.03$ count s$^{-1}$ (0.5 -- 10 keV). No previous detection of this source was made by S-CUBED which has been monitoring this region of the SMC approximately weekly since June 8th, 2016. Including this detection, \sw\ was detected 10 times between the start of the outburst and April 21st, 2020, and was not detected in 2 other observations taken during this period.

Starting April 14th, 2020, Swift began observing the source with deeper $\sim2$ks exposures every 1-2 days, in order to better characterize the spectrum and shorter term variability of the source, and determine a more accurate measurement of the spectrum. 

\begin{figure}
	\includegraphics[width=9cm,angle=-0]{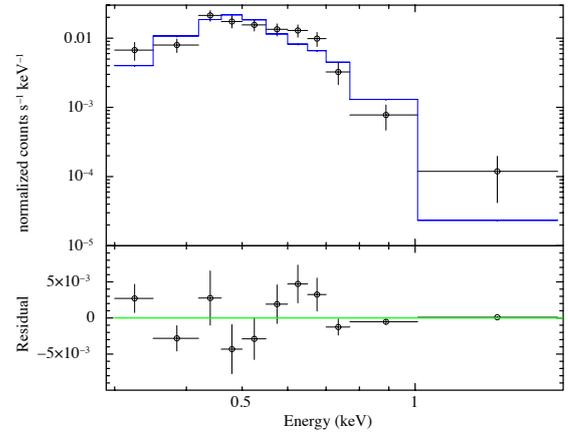}
    \caption{\textcolor{black}{Combined X-ray spectrum of \sw\ fit with an absorbed blackbody model. Note there is no significant X-ray emission above 2 keV.}}
    \label{fig:spectrum}
\end{figure}

The combined spectrum of the Swift data reveals that \sw\ is very soft, with no counts detected above 2 keV. Fitting a blackbody spectrum to the time averaged spectrum, and fixing the absorption column to the standard SMC value of $5.9 \times 10^{20} \mathrm{cm^{-2}}$ \citep{dl1990}, reveals a blackbody temperature of $kT = 90 \pm 6$~eV. \textcolor{black}{We note however that if we allow the absorption value to be a free parameter, the fit is significantly improved (reduced $\chi^2$ falls from $4.3$ to $1.5$) by having a higher absorption and lower temperature, $N_H = 3.4 \pm 2.0 \times 10^{21}\ \mathrm{cm^{-2}}$ and $kT = 58 \pm 6$~eV (see Figure \ref{fig:spectrum}). We note that although this higher absorption might be indicative of localized absorption, it is consistent within errors with the predicted X-ray absorption ($3.2 \times 10^{21}\ \mathrm{cm^{-2}}$) towards the source as derived using the method described in \cite{Willingale2013}, so we cannot rule out that the absorption is purely line-of-sight in origin.} This soft thermal spectrum would place \sw\ firmly in the category of Super Soft Sources, and \textcolor{black}{opens up the possibility} that the system contains a White Dwarf (WD) compact object.

\begin{figure}
	\includegraphics[width=9cm,angle=-0]{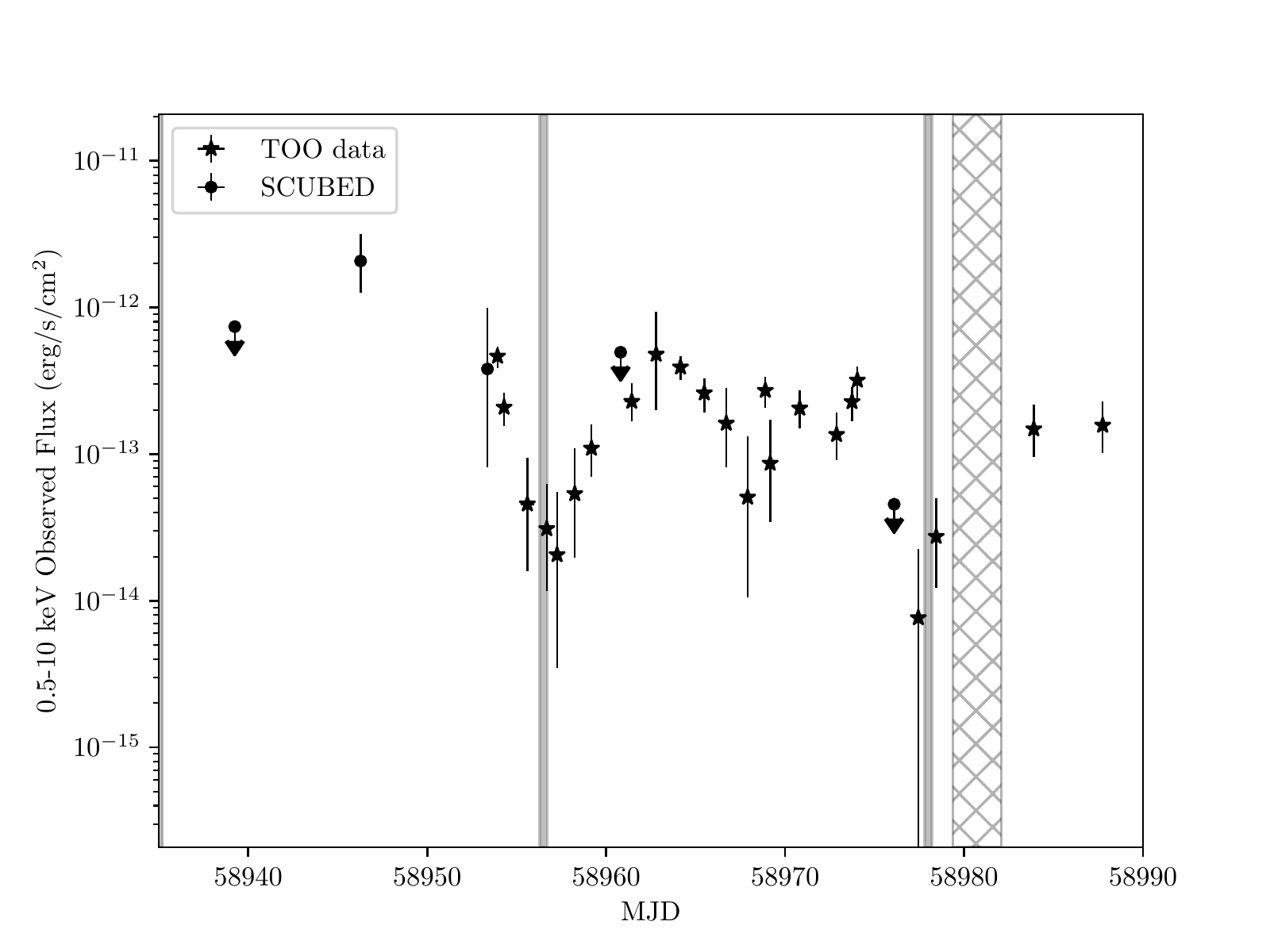}
    \caption{Light-curve of the outburst of \sw\ as seen by Swift/XRT. The vertical grey bands indicate the predicted time for the optical flare - see Section \ref{sec:ogle}. \textcolor{black}{The hatched region indicates a period when Swift could not observe the source due to an observation constraint.}}
    \label{fig:xray}
\end{figure}

\textcolor{black}
{We note that calculating a luminosity for this source is problematic, due to the relatively low quality spectrum leading to uncertainties the absorption, which means correcting for absorption will lead to large uncertainties in the luminosity values. Therefore} a light-curve of the \textcolor{black}{simply the observed flux, uncorrected for absorption,} for \sw\ was constructed utilizing data from both S-CUBED and the additional Swift monitoring, shown in Figure \ref{fig:xray}. 

The source is detected in S-CUBED analysis at a peak flux of $(2.1^{+0.8}_{-1.1}) \times 10^{-12}$~erg/s/cm$^2$ (0.5-10 keV), and then consistently after the initial observations. \textcolor{black}{Assuming a standard SMC distance of 62~kpc \citep{scowcroft2016}) and correcting for absorption fixed at the value derived from \cite{Willingale2013}, this translates into a 0.5-10~keV luminosity of $4.1^{1.6}_{-2.2} \times 10^{36}\ \mathrm{erg/s}$, with the uncertainties reflecting the uncertainties in the assumed absorption.} 

The higher cadence TOO observations show an apparent dipping structure, which shows the emission dropping significantly below the upper limits of individual S-CUBED non-detections. Although interpretation of this light-curve structure is not possible without longer monitoring, it is strongly suggestive of significant variability, probably linked to the 21.5 day optical period, \textcolor{black}{showing two apparent dips with the minima spacing consistent with the optical period}. 

No detection of \sw\ was made prior to its initial detection by S-CUBED in January 2020. By combining all of the S-CUBED data taken prior to the first detection, we calculate a $3\sigma$ upper-limit on the count rate of $2.1 \times 10^{-3}$~count~s$^{-1}$. Assuming the spectrum seen in outburst, this gives an upper limit on the source observed flux prior to detection of $4.4 \times 10^{-14}$~erg/s (0.5-10 keV). 
In addition, a catalogue search has found no previously detected X-ray source at this location. 

\subsection{The optical counterpart}

The best position of the X-ray source obtained from the Swift X-ray data is:

RA(2000)=00h 44m 27.98s, Dec(2000)=-73$^{\circ}$ 48' 04.3", error radius 2.6" (90\% confidence limit)\\

\begin{figure}

	\includegraphics[width=8cm,angle=-00]{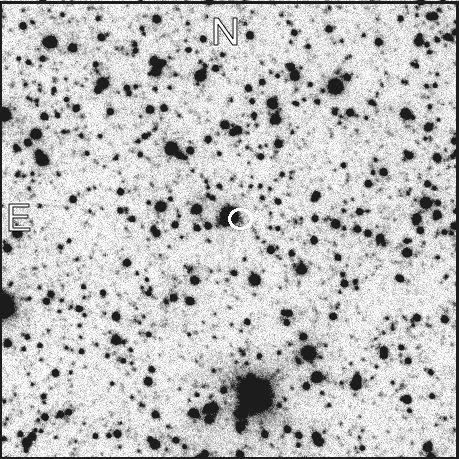}
    \caption{OGLE I band image of a 2 x 2 arcmin field centred on the optical counterpart. North is to the top, east to the left. The 90\% uncertainty of the position of \sw\ is shown by the circle.}
    \label{fig:fc}
\end{figure}

From Fig \ref{fig:fc} it can be seen that there is a bright stellar object with 1.5 arcseconds of the X-ray position, well within the uncertainties on that position. This optical counterpart to \sw ~is to be found in the catalogues of ~\cite{massey2002} and 2MASS. In the former it is identified as [M2002] 4294 and in the latter as 2MASS J00442806-7348031. The apparent magnitudes of the source in all available reported bands are shown in Table \ref{tab:phot}.

\begin{table}
    \centering
    \begin{tabular}{ccc}
        Photometric band&Mag&Error on mag\\
         \hline
         U&14.05&0.03\\
         B&14.96&0.03 \\
         V&15.00&0.03\\
         R&15.09&0.03\\
         I&14.95&0.01\\
         J&14.85&0.03\\
         H&14.81&0.04\\
         K&14.77&0.06\\
    \end{tabular}
    \caption{Photometric magnitudes of the optical counterpart to \sw.}
    \label{tab:phot}
\end{table}

\subsection{OGLE}
\label{sec:ogle}

\begin{figure*}

	\includegraphics[width=14cm,angle=-00]{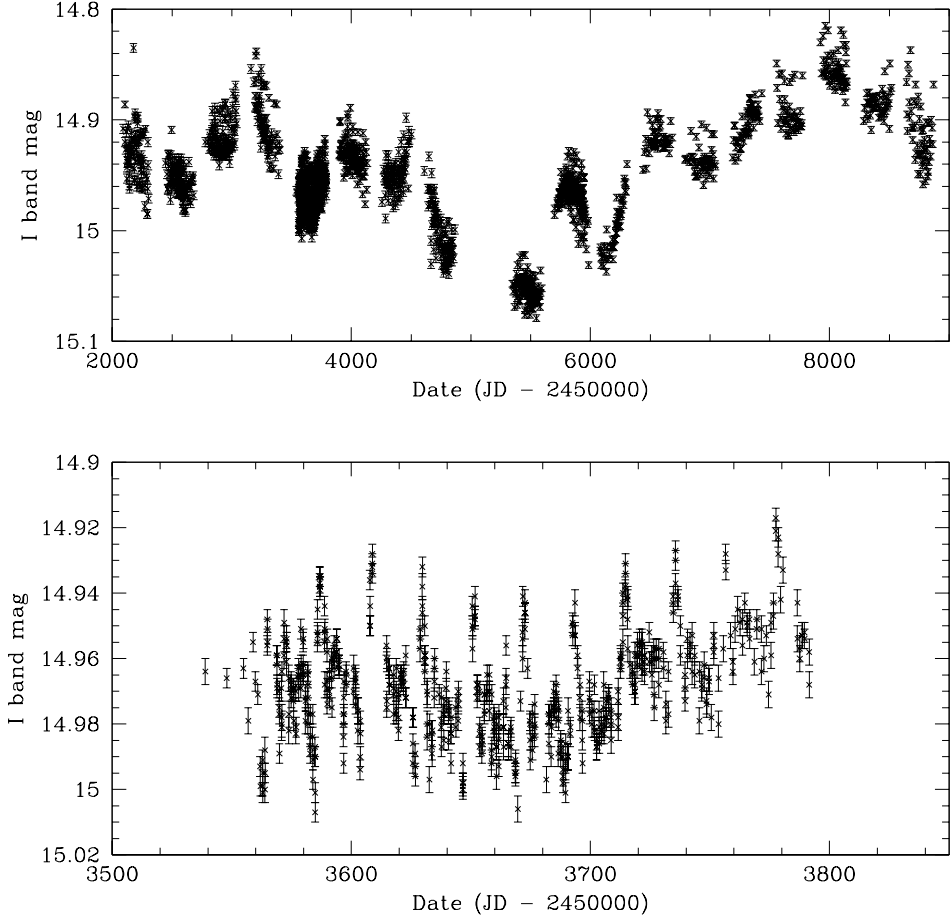}
    \caption{Upper panel - All the OGLE I-band data from the last$\sim$17 years. Lower panel - one year's worth of data.}
    \label{fig:lc2}
\end{figure*}

The OGLE project \citep{Udalski2015}  provides long term I-band photometry with a cadence of 1-3 days. it was possible to retrieve many years of photometric monitoring from OGLE III \& IV in the I-band, plus sparser coverage in the V band. The source is identified within the OGLE III project as SMC 720.13.12306 and as SMC 128.3.12469 in OGLE IV. All available data at the time of writing are included in this paper i.e. up to MJD 58869 (20 Jan 2020).

All the OGLE I-band data are presented in the upper panel of Fig \ref{fig:lc2}. 
Attempts to find a single coherent period for the flaring seen occasionally in the data (see lower panel of Fig \ref{fig:lc2}) were not successful. Instead the source exhibits periods of almost regular flaring, mostly at times when the source is in a brighter state - see Fig \ref{fig:lc2}. During these active periods the flaring frequency is almost stable and an ephemeris can be found. Most recently in the epoch MJD 57000-58869 the coherence is very strong and the outburst times of these flares is described by:\\

T$_{f}$ = 57000.0 + n(21.50$\pm$ 0.01) MJD \\

\begin{figure}

	\includegraphics[width=6cm,angle=-90]{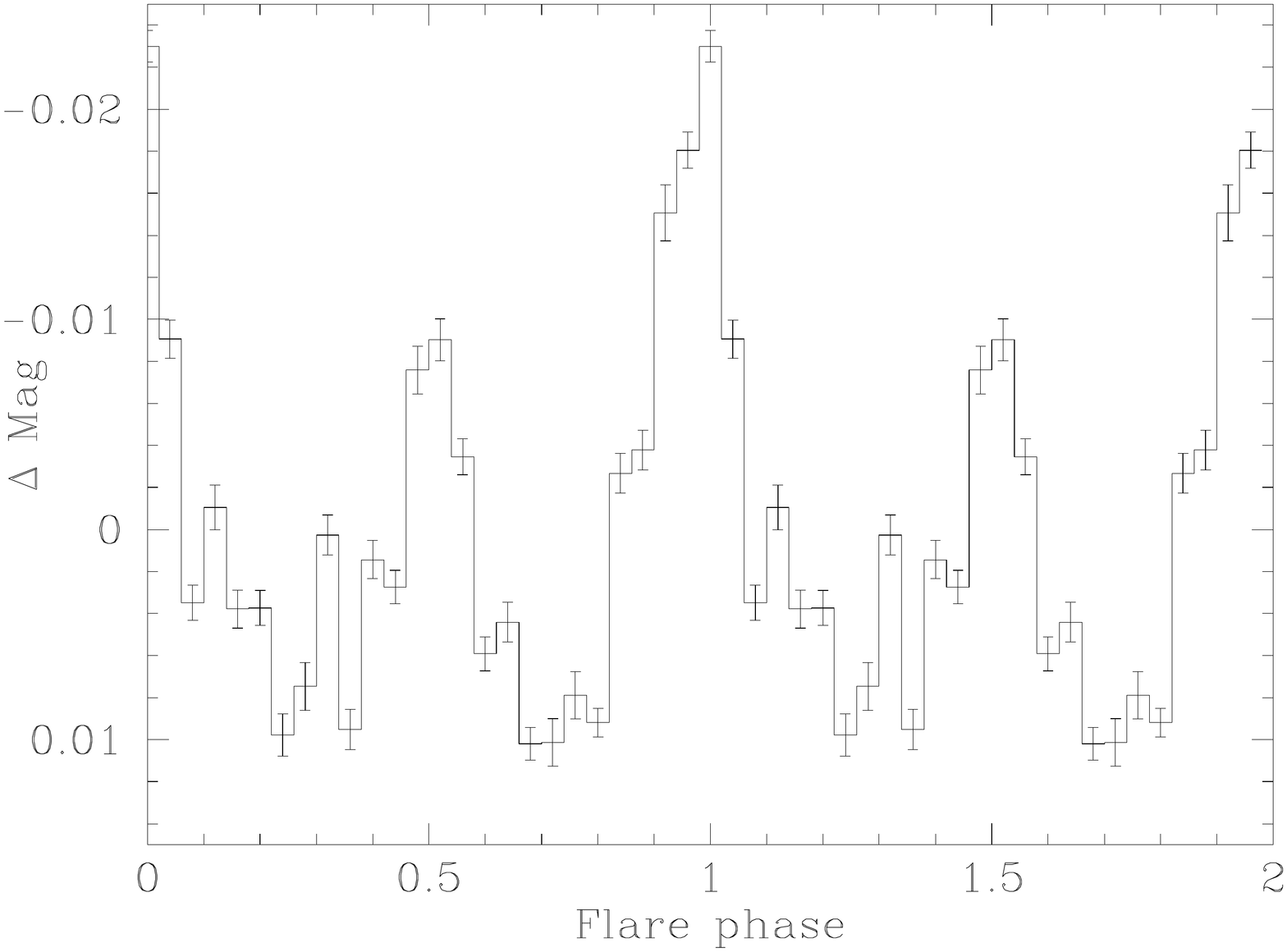}
    \caption{OGLE data for the period MJD 57000 - 58869 de-trended and then folded at the period of 21.50d. Phase 0.0 corresponds to the time of the optical flare.}
    \label{fig:fold}
\end{figure}

Data from this epoch were de-trended using a simple polynomial and then folded using this ephemeris. The result is shown in Fig \ref{fig:fold}. This figure reveals not only a clear flare around phase 0.0, but also evidence for a smaller flare close to phase 0.5.

\begin{figure}

	\includegraphics[width=8cm,angle=-00]{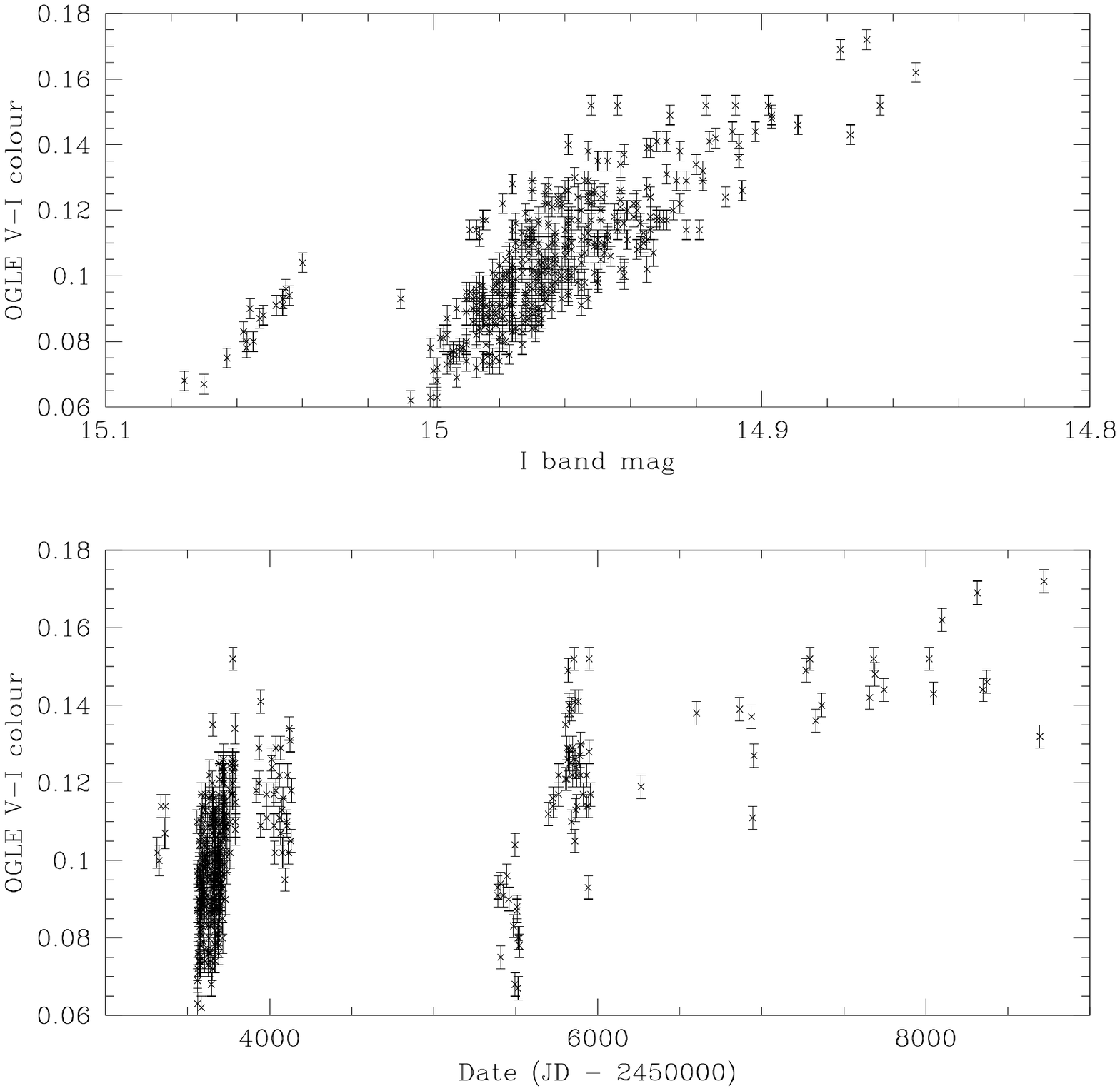}
    \caption{Colour changes in the OGLE data. Upper panel- (V-I) colour versus I band magnitude; lower panel - (V-I) colour versus time.}
    \label{fig:colour}
\end{figure}

Complicated colour changes are seen in the OGLE data and are presented in Fig \ref{fig:colour} and discussed below.

\section{Discussion}

\subsection{Classifying the optical counterpart to \sw}

\textcolor{black}{In the absence of a blue spectrum it is possible to make an estimate of the spectral class based on the observed magnitudes quoted in Table \ref{tab:phot} - assuming that the object lies in the SMC. Assuming the extinction to this object in the SMC is E(B-V)=0.07 \citep{skowron2020}  and a distance modules of 18.95 \citep{graczyk2014} it is possible to estimate the absolute V band magnitude of the star to be -4.17$\pm$0.03. From the tables in \cite{jj1987J} this falls between that of a O9 star (M$_{V}$ = -4.5) and a B0 star (M$_{V}$ = -4.0) for luminosity class V. And between B1 (M$_{V}$ = -4.4) and B2(M$_{V}$ = -3.9) if the luminosity class is III.}

\begin{figure}
	\includegraphics[width=8cm,angle=-0]{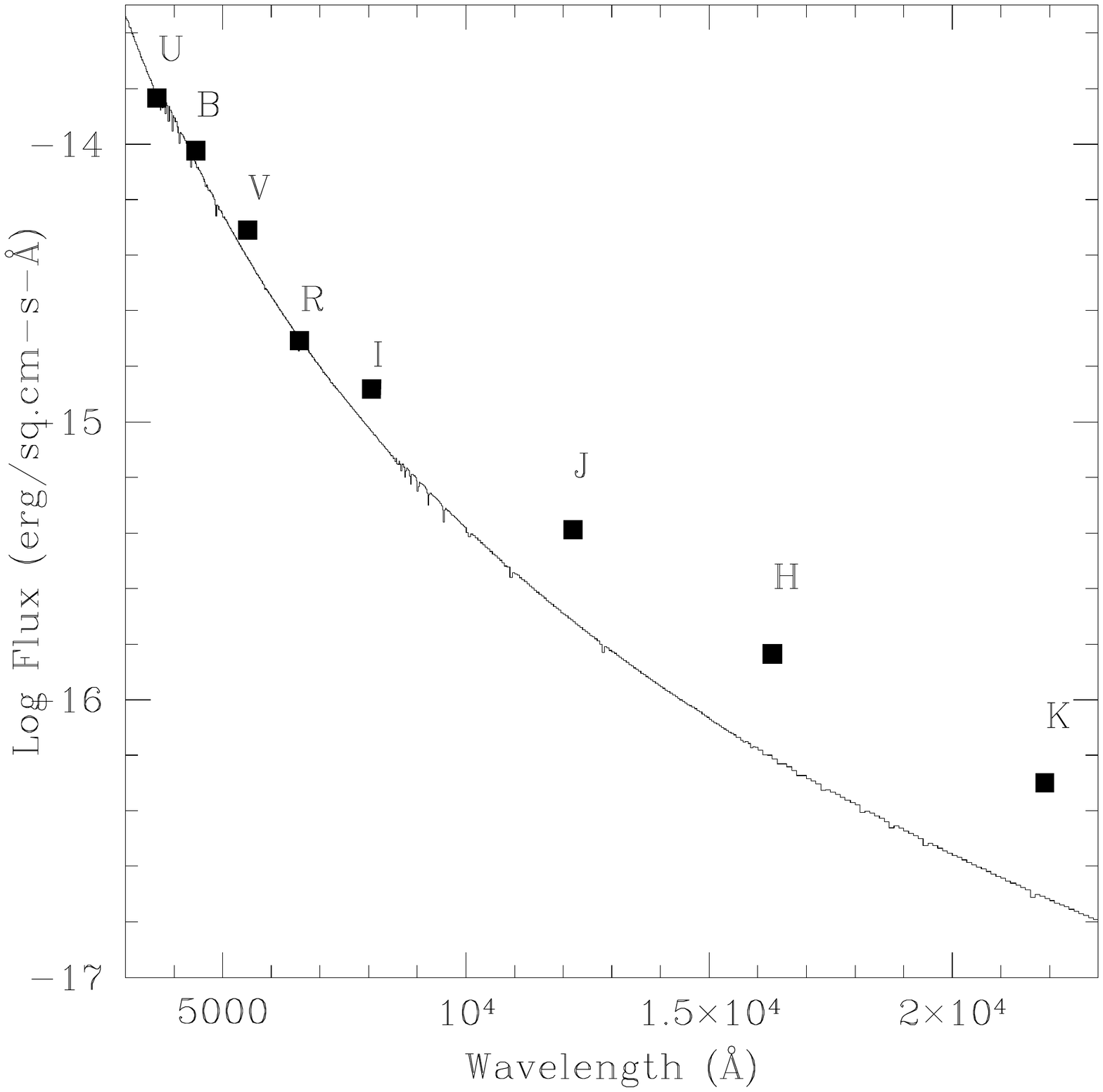}
    \caption{Comparison of the dereddened photometric measurements listed in Table \ref{tab:phot} with a Kurucz model for a O9.5V type star.}
    \label{fig:photom}
\end{figure}

Using the data in Table \ref{tab:phot} and correcting for the reddening to the SMC of E(B-V)=0.07 \citep{skowron2020} the measured photometry may be compared to a standard Kurucz model \citep{kurucz1979} for an O9.5V type star. This comparison is shown in Fig~\ref{fig:photom} where the model has been normalised to the U band data. From this figure it is very clear that this star shows a substantial red excess, \textcolor{black}{possibly}indicative of the presence of a circumstellar disk with an average lower temperature than the star itself.

\textcolor{black}{Including the possible evidence }for the clear presence of a circumstellar disk implies the best classification at the moment in the absence of a spectrum to be in the range between O9Ve and B2IIIe. Once a blue spectrum is obtained then a more explicit comparisons can be made with stellar models such as that done by \cite{orio2010} for the source in M31 or \cite{coe2020} for the source in the Magellanic Bridge.

\textcolor{black} {Finally we note that we can confidently rule out the possibility that this optical object is a foreground star. Its colours, location and magnitudes place it firmly amongst the population of OB stars in the SMC \citep{kourniotis2014}. Thus the strong link between the optical and X-ray behaviour (see Fig Fig \ref{fig:xray}) confirms that the X-ray source must also lie in the SMC.} 

\subsection{Optical \& X-ray variability}

The range in (V-I) colour in Fig.~\ref{fig:colour} is on the order of $\sim$0.1~mag, which is typically seen during disc-growth and decline in other Be/X-ray systems (see \citealt{Rajoelimanana2011}). This change in brightness is traditionally associated with the change in extent of a circumstellar disk around the Be star. This hypothesis is supported by the colour changes that are associated with the brightness variations. Fig \ref{fig:colour} was constructed from the occasions on which OGLE III \& IV obtained V \& I band data within 2 days of each other. In the top panel of this figure it can be seen that the large majority of the data points follow the pattern of increasing redness with increasing brightness. This would be totally consistent with the growth of a disk with an average temperature less than the $\sim$30,000K of the mass donor star. The lower panel of Fig \ref{fig:colour} shows the colour changes as a function of time and though the data are much sparser in recent times, it can be seen that the colours have been at the most red since $\sim$ MJD58000. Thus we can understand the recent detections of X-rays from this system in terms of the circumstellar disk having expanded far enough to get close enough to the orbital path of the compact object to provide fuel for accretion.

This evidence for the more intimate interaction between the compact object and the disk also provides an explanation for the emergence of the persistent optical flares shown in Fig \ref{fig:fold}. It is well established from Smooth Particle Hydrodynamic Simulations (SPH) of such systems \citep{okazaki2002} that the proximity of the compact object will cause the disk to distend around periastron, thereby increasing the surface area. Since these disks are optically thick this will naturally increase the total emission from the disk. So it is extremely likely that this flaring period of $\sim$21.5d must be closely linked to the binary period. The small variations observed in the exact phase of the flare arise from the slightly different points at which the compact object starts to interact with the disk - the larger the disk, the earlier the start of the flare. Probably the width of the outbursts will also be related to the circumstellar disk size.

It is also worth noting in Fig \ref{fig:fold} a significant suggestion of a smaller regular flare at Phase 0.5 - which we interpret as apastron for the compact object. This could represent an important constraint on the orbital solution for the system, suggesting the compact object has to approach the disk twice an orbit, once more closely than the other.

It is interesting to note the smaller colour track evident in the lower left hand corner of the upper panel of Fig \ref{fig:colour}. These are data recorded whilst the system is at its most faintest and it suggests that an alternative small disk may sometimes occur which fails to grow into the normal much larger system.

In order to try and understand why an X-ray dip is observed coincident with the predicted time of an optical flare some thought has to be given to the orbital configuration of the system.

\begin{figure}
	\includegraphics[width=7cm,angle=-0]{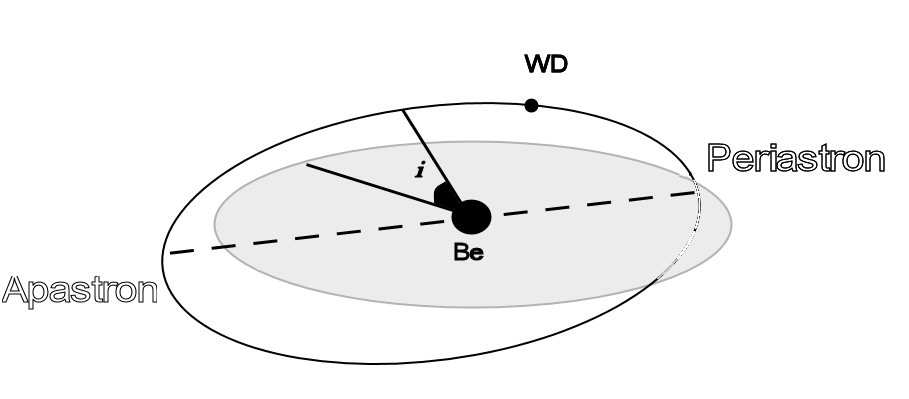}
    \caption{Illustration of the possible system configuration for \sw{}. The angle {\it i} is the angle between the plane of the orbit of the white dwarf and that of the circumstellar disk.}
    \label{fig:disk3}
\end{figure}

If the orbit of the white dwarf (WD) is inclined with respect to the circumstellar (CS) disk plane by an angle {\it i}, then there are two points when the WD will pass through the plane of the CS disk (not necessarily through the actual CS disk though). See Fig \ref{fig:disk3}. If one of these points is required to be at periastron, then the second point would naturally be at apastron. If the eccentricity is zero (circular orbit) then the two interactions would be identical in distance between the WD orbital path and the CS disk. But once you introduce some eccentricity then one will be a closer encounter (periastron) and one less-close encounter (apastron). That might explain a larger flare at phase 0.0 and a smaller flare at phase 0.5.

The observed dip in the X-ray flux is then understood as the obscuration/occultation of the WD by the edge of the CS disk around the time of periastron. It is possible with more accurate modelling to make a prediction for both the inclination angle {\it i} and the orbital eccentricity. The relative sizes of the two flares seen in the optical give clues to the magnitude of the eccentricity. Detailed SPH modelling of this system - as carried out for other Be/X-ray systems (see \cite{brown2019} for example) - could greatly help in our understanding of the observed behaviour reported here.

\section{Conclusions}

Reported in this work is \textcolor{black}{ one of the few systems proposed as a WD with a massive OB star companion in the SMC}. \textcolor{black}{If the optical counterpart is above 8 $M_\odot$ as the photometric classification suggests, and the compact object is indeed a white dwarf, this binary must have undergone an extreme mass transfer phenomenon.} Since binary evolution models predict large numbers of such systems compared to, for example, BeNS systems it is crucial to try and identify any candidates. There are $\sim$100 High Mass X-ray Binary systems catalogued in the SMC, and, so far, only \textcolor{black}{four proposed} BeWD ones. The detection of this \textcolor{black}{system could, therefore, be} very important. Furthermore, this particular system reveals complicated and intriguing time variability that should provide valuable inputs into understanding the orbital configuration, and the manner in which binary partners can disrupt a circumstellar disk structure. 

\section*{Acknowledgements}

The OGLE project has received funding from the National Science Centre,
Poland, grant MAESTRO 2014/14/A/ST9/00121 to AU. This publication makes use of data products from the Two Micron All Sky Survey, which is a joint project of the University of Massachusetts and the Infrared Processing and Analysis Center/California Institute of Technology, funded by the National Aeronautics and Space Administration and the National Science Foundation. PAE acknowledges UKSA support. Mark Sullivan is acknowledged for his help in making the data analysis for this paper possible. An anonymous referee is thanked for their helpful insights into the context of this discovery.




\bibliographystyle{mnras}
\bibliography{references}

\begin{thebibliography}{}
\makeatletter
\relax
\def\mn@urlcharsother{\let\do\@makeother \do\$\do\&\do\#\do\^\do\_\do\%\do\~}
\def\mn@doi{\begingroup\mn@urlcharsother \@ifnextchar [ {\mn@doi@}
  {\mn@doi@[]}}
\def\mn@doi@[#1]#2{\def\@tempa{#1}\ifx\@tempa\@empty \href
  {http://dx.doi.org/#2} {doi:#2}\else \href {http://dx.doi.org/#2} {#1}\fi
  \endgroup}
\def\mn@eprint#1#2{\mn@eprint@#1:#2::\@nil}
\def\mn@eprint@arXiv#1{\href {http://arxiv.org/abs/#1} {{\tt arXiv:#1}}}
\def\mn@eprint@dblp#1{\href {http://dblp.uni-trier.de/rec/bibtex/#1.xml}
  {dblp:#1}}
\def\mn@eprint@#1:#2:#3:#4\@nil{\def\@tempa {#1}\def\@tempb {#2}\def\@tempc
  {#3}\ifx \@tempc \@empty \let \@tempc \@tempb \let \@tempb \@tempa \fi \ifx
  \@tempb \@empty \def\@tempb {arXiv}\fi \@ifundefined
  {mn@eprint@\@tempb}{\@tempb:\@tempc}{\expandafter \expandafter \csname
  mn@eprint@\@tempb\endcsname \expandafter{\@tempc}}}

\bibitem[\protect\citeauthoryear{{Brown}, {Coe}, {Ho}  \& {Okazaki}}{{Brown}
  et~al.}{2019}]{brown2019}
{Brown} R.~O.,  {Coe} M.~J.,  {Ho} W.~C.~G.,   {Okazaki} A.~T.,  2019, \mn@doi
  [\mnras] {10.1093/mnras/stz1037}, \href
  {https://ui.adsabs.harvard.edu/abs/2019MNRAS.486.3078B} {486, 3078}

\bibitem[\protect\citeauthoryear{{Burrows} et~al.,}{{Burrows}
  et~al.}{2005}]{burrows05}
{Burrows} D.~N.,  et~al., 2005, \mn@doi [\ssr] {10.1007/s11214-005-5097-2},
  \href {http://adsabs.harvard.edu/abs/2005SSRv..120..165B} {120, 165}

\bibitem[\protect\citeauthoryear{{Coe} \& {Kirk}}{{Coe} \&
  {Kirk}}{2015}]{ck2015}
{Coe} M.~J.,  {Kirk} J.,  2015, \mn@doi [\mnras] {10.1093/mnras/stv1283}, \href
  {https://ui.adsabs.harvard.edu/abs/2015MNRAS.452..969C} {452, 969}

\bibitem[\protect\citeauthoryear{{Coe}, {Kennea}, {Evans}  \& {Udalski}}{{Coe}
  et~al.}{2020}]{coe2020}
{Coe} M.~J.,  {Kennea} J.~A.,  {Evans} P.,   {Udalski} A.,  2020, The
  Astronomer's Telegram, \href
  {https://ui.adsabs.harvard.edu/abs/2020ATel13626....1C} {13626, 1}

\bibitem[\protect\citeauthoryear{{Cracco}, {Orio}, {Ciroi}, {Gallagher},
  {Kotulla}  \& {Romero-Colmenero}}{{Cracco} et~al.}{2018}]{cracco2018}
{Cracco} V.,  {Orio} M.,  {Ciroi} S.,  {Gallagher} J.,  {Kotulla} R.,
  {Romero-Colmenero} E.,  2018, \mn@doi [\apj] {10.3847/1538-4357/aacefb},
  \href {https://ui.adsabs.harvard.edu/abs/2018ApJ...862..167C} {862, 167}

\bibitem[\protect\citeauthoryear{{Dickey} \& {Lockman}}{{Dickey} \&
  {Lockman}}{1990}]{dl1990}
{Dickey} J.~M.,  {Lockman} F.~J.,  1990, \mn@doi [\araa]
  {10.1146/annurev.aa.28.090190.001243}, \href
  {https://ui.adsabs.harvard.edu/abs/1990ARA&A..28..215D} {28, 215}

\bibitem[\protect\citeauthoryear{{Gehrels} et~al.,}{{Gehrels}
  et~al.}{2004}]{gehrels04}
{Gehrels} N.,  et~al., 2004, \mn@doi [\apj] {10.1086/422091}, \href
  {https://ui.adsabs.harvard.edu/abs/2004ApJ...611.1005G} {611, 1005}

\bibitem[\protect\citeauthoryear{{Graczyk} et~al.,}{{Graczyk}
  et~al.}{2014}]{graczyk2014}
{Graczyk} D.,  et~al., 2014, \mn@doi [\apj] {10.1088/0004-637X/780/1/59}, \href
  {https://ui.adsabs.harvard.edu/abs/2014ApJ...780...59G} {780, 59}

\bibitem[\protect\citeauthoryear{{Haberl} et~al.,}{{Haberl}
  et~al.}{2012}]{haberl2012}
{Haberl} F.,  et~al., 2012, \mn@doi [\aap] {10.1051/0004-6361/201219758}, 545,
  A128

\bibitem[\protect\citeauthoryear{{Jaschek} \& {Jaschek}}{{Jaschek} \&
  {Jaschek}}{1987}]{jj1987J}
{Jaschek} C.,  {Jaschek} M.,  1987, {The classification of stars}.
Cambridge University Press

\bibitem[\protect\citeauthoryear{{Kahabka}}{{Kahabka}}{2006}]{kahabka2006}
{Kahabka} P.,  2006, \mn@doi [Advances in Space Research]
  {10.1016/j.asr.2005.10.058}, \href
  {https://ui.adsabs.harvard.edu/abs/2006AdSpR..38.2836K} {38, 2836}

\bibitem[\protect\citeauthoryear{{Kahabka}, {Haberl}, {Payne}  \&
  {Filipovi{\'c}}}{{Kahabka} et~al.}{2006}]{k2006}
{Kahabka} P.,  {Haberl} F.,  {Payne} J.~L.,   {Filipovi{\'c}} M.~D.,  2006,
  \mn@doi [\aap] {10.1051/0004-6361:20065490}, \href
  {https://ui.adsabs.harvard.edu/abs/2006A&A...458..285K} {458, 285}

\bibitem[\protect\citeauthoryear{{Kennea}, {Coe}, {Evans}, {Waters}  \&
  {Jasko}}{{Kennea} et~al.}{2018}]{kennea2018}
{Kennea} J.~A.,  {Coe} M.~J.,  {Evans} P.~A.,  {Waters} J.,   {Jasko} R.~E.,
  2018, \mn@doi [\apj] {10.3847/1538-4357/aae839}, \href
  {https://ui.adsabs.harvard.edu/abs/2018ApJ...868...47K} {868, 47}

\bibitem[\protect\citeauthoryear{{Kourniotis} et~al.,}{{Kourniotis}
  et~al.}{2014}]{kourniotis2014}
{Kourniotis} M.,  et~al., 2014, \mn@doi [\aap] {10.1051/0004-6361/201322856},
  \href {https://ui.adsabs.harvard.edu/abs/2014A&A...562A.125K} {562, A125}

\bibitem[\protect\citeauthoryear{{Kurucz}}{{Kurucz}}{1979}]{kurucz1979}
{Kurucz} R.~L.,  1979, \mn@doi [\apjs] {10.1086/190589}, \href
  {https://ui.adsabs.harvard.edu/abs/1979ApJS...40....1K} {40, 1}

\bibitem[\protect\citeauthoryear{{Massey}}{{Massey}}{2002}]{massey2002}
{Massey} P.,  2002, \mn@doi [\apjs] {10.1086/338286}, \href
  {https://ui.adsabs.harvard.edu/abs/2002ApJS..141...81M} {141, 81}

\bibitem[\protect\citeauthoryear{{Okazaki}, {Bate}, {Ogilvie}  \&
  {Pringle}}{{Okazaki} et~al.}{2002}]{okazaki2002}
{Okazaki} A.~T.,  {Bate} M.~R.,  {Ogilvie} G.~I.,   {Pringle} J.~E.,  2002,
  \mn@doi [\mnras] {10.1046/j.1365-8711.2002.05960.x}, \href
  {https://ui.adsabs.harvard.edu/abs/2002MNRAS.337..967O} {337, 967}

\bibitem[\protect\citeauthoryear{{Oliveira}, {Steiner}, {Ricci}, {Menezes}  \&
  {Borges}}{{Oliveira} et~al.}{2010}]{oliveira2010}
{Oliveira} A.~S.,  {Steiner} J.~E.,  {Ricci} T.~V.,  {Menezes} R.~B.,
  {Borges} B.~W.,  2010, \mn@doi [\aap] {10.1051/0004-6361/201014773}, \href
  {https://ui.adsabs.harvard.edu/abs/2010A&A...517L...5O} {517, L5}

\bibitem[\protect\citeauthoryear{{Orio}, {Nelson}, {Bianchini}, {Di Mille}  \&
  {Harbeck}}{{Orio} et~al.}{2010}]{orio2010}
{Orio} M.,  {Nelson} T.,  {Bianchini} A.,  {Di Mille} F.,   {Harbeck} D.,
  2010, \mn@doi [\apj] {10.1088/0004-637X/717/2/739}, \href
  {https://ui.adsabs.harvard.edu/abs/2010ApJ...717..739O} {717, 739}

\bibitem[\protect\citeauthoryear{{Raguzova}}{{Raguzova}}{2001}]{raguzova2001}
{Raguzova} N.~V.,  2001, \mn@doi [\aap] {10.1051/0004-6361:20000348}, \href
  {https://ui.adsabs.harvard.edu/abs/2001A&A...367..848R} {367, 848}

\bibitem[\protect\citeauthoryear{{Rajoelimanana}, {Charles}  \&
  {Udalski}}{{Rajoelimanana} et~al.}{2011}]{Rajoelimanana2011}
{Rajoelimanana} A.~F.,  {Charles} P.~A.,   {Udalski} A.,  2011, \mn@doi
  [\mnras] {10.1111/j.1365-2966.2011.18243.x}, \href
  {https://ui.adsabs.harvard.edu/abs/2011MNRAS.413.1600R} {413, 1600}

\bibitem[\protect\citeauthoryear{{Scowcroft}, {Freedman}, {Madore}, {Monson},
  {Persson}, {Rich}, {Seibert}  \& {Rigby}}{{Scowcroft}
  et~al.}{2016}]{scowcroft2016}
{Scowcroft} V.,  {Freedman} W.~L.,  {Madore} B.~F.,  {Monson} A.,  {Persson}
  S.~E.,  {Rich} J.,  {Seibert} M.,   {Rigby} J.~R.,  2016, \mn@doi [\apj]
  {10.3847/0004-637X/816/2/49}, \href
  {https://ui.adsabs.harvard.edu/abs/2016ApJ...816...49S} {816, 49}

\bibitem[\protect\citeauthoryear{{Skowron et al.}}{{Skowron et
  al.}}{2020}]{skowron2020}
{Skowron et al.} 2020, In prep.

\bibitem[\protect\citeauthoryear{{Sturm}, {Haberl}, {Pietsch}, {Coe},
  {Mereghetti}, {La Palombara}, {Owen}  \& {Udalski}}{{Sturm}
  et~al.}{2012}]{sturm2012}
{Sturm} R.,  {Haberl} F.,  {Pietsch} W.,  {Coe} M.~J.,  {Mereghetti} S.,  {La
  Palombara} N.,  {Owen} R.~A.,   {Udalski} A.,  2012, \mn@doi [\aap]
  {10.1051/0004-6361/201117789}, \href
  {https://ui.adsabs.harvard.edu/abs/2012A&A...537A..76S} {537, A76}

\bibitem[\protect\citeauthoryear{{Udalski}, {Szyma{\'n}ski}  \&
  {Szyma{\'n}ski}}{{Udalski} et~al.}{2015}]{Udalski2015}
{Udalski} A.,  {Szyma{\'n}ski} M.~K.,   {Szyma{\'n}ski} G.,  2015, \actaa,
  \href {https://ui.adsabs.harvard.edu/abs/2015AcA....65....1U} {65, 1}

\bibitem[\protect\citeauthoryear{{Willingale}, {Starling}, {Beardmore},
  {Tanvir}  \& {O'Brien}}{{Willingale} et~al.}{2013}]{Willingale2013}
{Willingale} R.,  {Starling} R.~L.~C.,  {Beardmore} A.~P.,  {Tanvir} N.~R.,
  {O'Brien} P.~T.,  2013, \mn@doi [\mnras] {10.1093/mnras/stt175}, \href
  {https://ui.adsabs.harvard.edu/abs/2013MNRAS.431..394W} {431, 394}

\makeatother
\end{thebibliography}

\label{lastpage}
\bsp	

\end{document}